\documentclass[runningheads]{llncs}
\usepackage{graphicx}
\usepackage{hhline}

\usepackage{amssymb}
\usepackage{amsmath}
\usepackage{bbm}

\usepackage{caption}
\usepackage{subcaption}
\usepackage{booktabs}
\usepackage{tabularx}
\usepackage{longtable}
\usepackage{xltabular}

\begin{document}
\title{Improving Phenotype Prediction using Long-Range Spatio-Temporal Dynamics of Functional Connectivity}
\titlerunning{Brain-MS-G3D}

\author{Simon Dahan\inst{1} \and Logan Z. J. Williams\inst{1,2} \and Daniel Rueckert \inst{3,4} \and Emma C. Robinson\inst{1,2}}


\authorrunning{S. Dahan et al.}
%
\institute{Department of Biomedical Engineering, School of Biomedical Engineering and Imaging Sciences, King’s College London, London, SE1 7EH, UK  \and Centre for the Developing Brain, Department of Perinatal Imaging and Health, School of Biomedical Engineering and Imaging Sciences, King’s College London, London, SE1 7EH, UK  \and Biomedical Image Analysis Group, Department of Computing, Imperial College London, London, SW7 2AZ, UK \and Klinikum rechts der Isar, Technical University of Munich, Munich, 81675, Germany}
\maketitle              
%
\begin{abstract}

The study of functional brain connectivity (FC) is important for understanding the underlying mechanisms of many psychiatric disorders.
Many recent analyses adopt graph convolutional networks, to study non-linear interactions between functionally-correlated states. However, although patterns of brain activation are known to be hierarchically organised in both \textit{space and time}, many methods have failed to extract powerful spatio-temporal features.
To overcome those challenges, and improve understanding of long-range functional dynamics, we translate an approach, from the domain of skeleton-based action recognition, designed to model interactions across space and time. 
We evaluate this approach using the Human Connectome Project (HCP) dataset on sex classification and fluid intelligence prediction. To account for subject topographic variability of functional organisation, we modelled functional connectomes using multi-resolution  dual-regressed (subject-specific) ICA nodes. Results show a prediction accuracy of 94.4\% for sex classification (an increase of 6.2\% compared to other methods), and an improvement of correlation with fluid intelligence of 0.325 vs 0.144, relative to a baseline model that encodes space and time separately. Results suggest that explicit encoding of spatio-temporal dynamics of brain functional activity may improve the precision with which behavioural and cognitive phenotypes may be predicted in the future.

\keywords{Human Connectome Project  \and Graph Convolution Networks \and Temporal Dynamics \and Functional MRI \and Phenotyping.}
\end{abstract}

\section*{Introduction}
Functional MRI (fMRI) is widely recognised as a cornerstone technique for relating brain processes to behaviour \cite{S.Smith2013,R.Liegeois2019}. In particular, the correlation between regional time-series activations, referred to as functional connectivity (FC), is known to be subject-specific \cite{E.Finn2015,R.Kong2019} and discriminate between populations \cite{S.Smith2015,S.Ktena2018}. This has led to it being popularly investigated as a potential clinical biomarker for many psychiatric and neurological disorders \cite{S.Ktena2018,Z.Huang2020}.
Despite several promising studies on large cohorts, for example, the Autism Spectrum Disorder study ABIDE \cite{S.Ktena2018}, several recent works have emphasised the need to better account for morphological, or topographic, heterogeneity of cortical architecture when building functional networks  \cite{J.Bijsterbosch2018,R.Kong2021,M.Glasser2016}. This is considered especially important for disease modelling \cite{A.Marquand2019}, where phenotypes are often heterogeneous and may be classified into sub-types \cite{T.Wolfers2021}. Under such circumstances fitting all data to a single global-average model of healthy functional organisation, stands to mask subtle features of disease \cite{A.Marquand2019,T.Wolfers2021}, as well as obscure understanding of mechanisms of disorders that usually co-occur \cite{T.Wolfers2021}. For these reasons, increasing efforts are being made to look towards individual subject-level analyses and move away from the case-control approach \cite{A.Marquand2016}.

To this end, recent approaches have looked towards better capture of inter-subject variation when deriving functional networks. For example, probabilistic functional modes (or PROFUMO) \cite{S.Harrison2020} uses a variational form of independent component analysis (ICA) to explicitly account for subject-specific variability during group-wise factorisation of fMRI data into temporal and spatial modes; \cite{R.Kong2019} implements a  multi-session hierarchical Bayesian model (MS-HBM) in order to cluster functional connectivity data into group-wise comparable, but subject-specific, functional parcellations; and \cite{M.Glasser2016} projects a hand-annotated, population-average multimodal parcellation of the cortex onto individuals by training a multi-layer perceptron (MLP) classifier; in the process revealing considerable topographic variation of regions when observed across the group.

At the same time, many studies are moving away from modelling static functional connectivity (averaged over all time) towards explicit modelling of temporal dynamics, as a means of better capturing temporal markers of behaviour \cite{D.Vidaurre2017,S.Gadgil2020,B.Kim2021,U.Pervaiz2021}. Of particular note, for this paper, are methods which translate models used for video understanding \cite{S.Yan2018},
especially graph convolution networks (GCN) used in skeleton-based action recognition \cite{S.Gadgil2020}. In this paper we specifically investigate the potential for translation of new techniques which explicitly encode interactions, across nodes, in both space and time \cite{Z.Liu2020}.
The specific contributions of this paper are as follows:
\begin{enumerate}
    \item \textbf{Explicit modelling of FC spatio-temporal dynamics improves prediction of behavioural phenotypes}: We demonstrate improvements over existing GCN models of dynamic FC \cite{S.Gadgil2020} by adopting a new model \cite{Z.Liu2020}, which aggregates functional signals in both space and time.
    \item \textbf{Accounting for inter-subject cortical heterogeneity improves performance}: We compare models trained using a 22 region group-average parcellation and multi-resolution dual-regressed (subject-specific) ICA nodes, and demonstrate gains in performance by accounting for topographic variability of functional organisation.
\end{enumerate}

\section*{Related Works}

The use of GCNs to study FC supports learning of complex, non-linear, long-range interactions between correlated functional network states. While early approaches focused on the analysis of static FC \cite{S.Ktena2018,X.Li2020,N.Dsouza2021}, demonstrating improved performance in metric learning for behavioural prediction of neuropsychiatric disorders such as Autism Spectrum Disorder \cite{S.Ktena2018,X.Li2020}; more recent papers have started to investigate the potential of GCNs to model FC temporal dynamics \cite{S.Gadgil2020,B.Kim2021}. Of these, \cite{S.Gadgil2020} takes inspiration from a skeleton-based GCN model for video understanding \cite{S.Yan2018}, translating this to the domain of dynamic FC modelling, by building a spatial graph from static FC (estimated across all node time series) but self-connecting nodes temporally along a regular grid. By applying spatial-only and temporal-only convolutions, the method successfully captured long-range (spatial and temporal) dependencies but failed to extract powerful space-time features by decoupling space and time graphs. By contrast, the more recent MS-G3D of \cite{Z.Liu2020} goes further to capture spatio-temporal semantics between nodes by encoding them through novel graph convolutional units which aggregate across space and time; in this way encoding complex actions, for example, the action of sitting down or standing up, in which the top half of a skeleton moves before the bottom. In this paper, we hypothesise that such long-range spatio-temporal behaviour is also observed in brain functional dynamics. We, therefore, benchmark this method against the method proposed by \cite{S.Gadgil2020} as a means for quantifying this effect.\footnote{The code for this experiment will be made available at: \url{http://www.github.com/metrics-lab/ST-fMRI/}} 

\section*{Methods}

\paragraph{\textbf{Data}}

\begin{figure}[h]
     \centering
     \includegraphics[width=12cm]{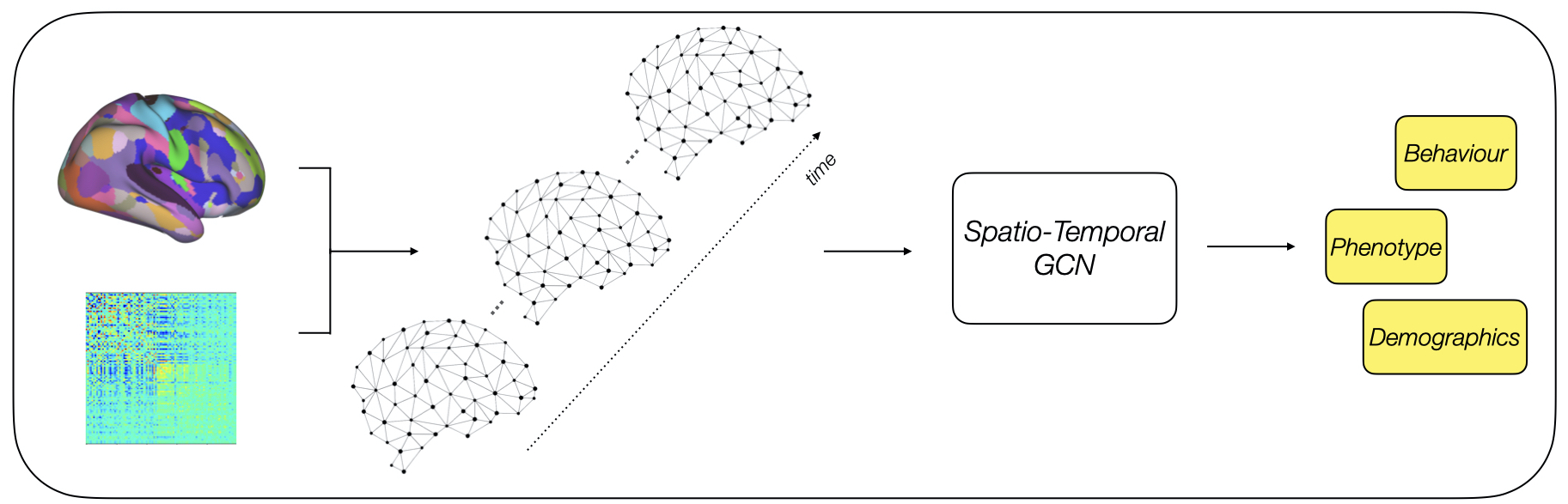}
     \caption{
     Illustration of the pipeline used in this study.
     Group-ICA parcellation regressed into individuals is combined with FC to build a spatio-temporal graph sequence representing time courses of brain activation.}
     \label{pipeline}
\end{figure}

This study used functional data from the first session (15min - 1200 frames - 0.72s/frame) of the Human Connectome Project (HCP) S1200 release \cite{D.VanEssen2013}, where individual subject node timeseries for 1003 participants were derived following group-wise independent component analysis (Group-ICA) \cite{C.Beckmann2004,S.Smith2013}. In brief, individual functional MRI data was processed through the HCP pipelines \cite{M.Glasser2013,S.Smith2013} and aligned using MSMAll \cite{E.Robinson2014,E.Robinson2018}. Then, spatial Group-ICA (with 468 subjects) was applied at different dimensionalities (15 to 300), factorising the data into a set of spatial maps (functional nodes) and associated time courses. These were propagated into individuals using dual regression, in this way tailoring the brain parcellations and time courses at the individual level (illustrated in figure \ref{pipeline}). We used all available data for the sex classification task (534 males and 469 females), but had to exclude 4 subjects for the fluid intelligence prediction task (due to missing data). We used Penn Progressive Matrices as scores of fluid intelligence, referred as \emph{PMAT24\_A\_CR} in the HCP behavioral data. Similarly to \cite{S.Gadgil2020}, we also benchmark against use of \textit{group-average} functional parcellation, where the group average HCP multimodal cortical parcellation (180 regions per hemisphere) \cite{M.Glasser2016} was aggregated into 22 regions of interest (ROIs) per hemisphere by combining nodes with common functions and large contiguous areas. In this latter configuration, nodes time series correspond to the average BOLD signal in each ROI. Subject-specific ICA nodes and group-averaged 22 ROIs time series were normalised by z-scores. 

\paragraph{\textbf{Model}}
In this paper, long-range, spatio-temporal dynamics of FC are modelled by extending the ST-GCN model \cite{S.Yan2018} (utilised in \cite{S.Gadgil2020}) to incorporate information over multiple spatial and temporal scales, by adapting the multi-scale, spatio-temporal graph operator (MS-G3D) approach of \cite{Z.Liu2020}.
Following the ST-GCN model of FC used by \cite{S.Gadgil2020}, functional connectomes are modelled as graphs $\mathcal{G}=(\mathcal{V},\mathcal{E})$, with $\mathcal{V}=\{v_1,v_2 .... v_N\}$ representing the set of functional regions (or nodes); and $\mathcal{E}$  representing spatial and temporal connections (or edges).
In this framework, the spatial graph \begin{math} \mathbf{A} \in \mathbb{R}^{\mathbb{N}\times\mathbb{N}} \end{math}:
\begin{equation}
    \mathbf{A}_{i, j}= \left\{\begin{array}{c} corr(i,j) \text { if } i\neq j \\0 \text { otherwise }\end{array}\right.
\end{equation}
represents static FC estimated over all time, with $corr(i,j)$ measuring the magnitude of correlation over the full timeseries (1200 frames). Spatio-temporal interactions are then modelled in the form of a feature tensor $ \mathbf{X} \in \mathbb{R}^{T \times N \times C}$:
\begin{math}\mathcal{X} = \allowbreak\left \{ x_{n,t} \in \mathbb{R}^{C} | t,n \in\mathbb{Z},\break 1\leq t\leq T, 1\leq n\leq N \right \} \end{math}, where the channel dimension of the input feature tensor represents the node time courses ($C=1$). 
The MS-G3D framework \cite{Z.Liu2020} models spatio-temporal graph convolutions (ST-GC) as blocks composed of two streams: a \emph{factorised pathway} which decouples space and time, and a \emph{G3D pathway} which aggregates space-time information. In \emph{the factorised pathway}, space-only convolutions are represented by:
\begin{equation} \mathbf{X}_{t}^{(l+1)} = \sigma\left ( \mathbf{\widetilde{D}}^{-1/2} (\mathbf{A}+\mathbf{I}) \mathbf{\widetilde{D}}^{-1/2} \mathbf{X}_{t}^{(l)} \Theta^{(l)}\right )
\label{gcn_conv}
\end{equation}
with $\Theta^{(l)}$ the learnable weights at layer $\textit{l}$ and $\mathbf{\widetilde{D}}$ the degree matrix of $(\mathbf{A}+\mathbf{I})$. Here, $\sigma$ is a ReLU activation function, except for the last ST-GC block of the model. Still in \emph{the factorised pathway}, temporal-only convolutions (TCN) are implemented with a bottleneck design, including four convolutional filters, all with kernels of size (3,1) along the temporal dimension but with various dilation factors ($d \in [1,4]$), and residual connections. In parallel, the \emph{G3D pathway} splits the full frame sequence into sub time-windows of $\tau \in \left \{ 3,5 \right \}$, and combines regional spacetime information using a bespoke spatio-temporal graph convolution operation (G3D), applied on every single sub time-window:
\begin{equation}
    \left [  \mathbf{X}_{(\tau)}^{(l+1)} \right ]_{t} = \sigma\left ( \mathbf{\widetilde{D}}^{-1/2}_{(\tau)} \mathbf{\tilde{A}}_{(\tau)} 
 \mathbf{\widetilde{D}}^{-1/2}_{(\tau)} \left[   \mathbf{X}_{(\tau)}^{(l)}\right ]_{t} \Theta^{(l)}\right )
\label{gctn_conv}
\end{equation}

with 
\begin{equation}
 \mathbf{\tilde{A}}_{(\tau)}=\left[\begin{array}{ccc}
\mathbf{(A+I)} & \ldots & \mathbf{(A+I)} \\
\vdots & \ddots & \vdots \\
\mathbf{(A+I)} & \cdots & \mathbf{(A+I)}
\end{array}\right] \in \mathbb{R}^{\mathbb{\tau N}\times\mathbb{\tau N}}
\label{stack}
\end{equation}
The G3D operation (Eq (\ref{gctn_conv})), therefore extends the spatial convolution (Eq (\ref{gcn_conv})) into a space-time operation by stacking the adjacency matrix $\mathbf{(A+I)}$ into $\mathbf{\tilde{A}}_{(\tau)}$ connecting nodes across $\tau$ frames, in space and time: edges being extended to connect nodes to their neighbours across the entire sub time-window; in this way constructing a spatio-temporal subgraph $\mathcal{G}_{(t)}=(\mathcal{V}_{(t)},\mathcal{E}_{(t)})$. $\mathbf{\widetilde{D}}_{(\tau)}$ and $\mathbf{X}_{(\tau)}$ are natural extensions of the previously introduced notations. 

In addition, a multi-scale (MS) operator can be embedded into any of the previously described convolutions. This aims at aggregating more structural information, by relating a node to its K-closest neighbours, through substituting the convolution operation in (Eq \ref{gcn_conv}) by  $\sum_{k=0}^{K} \mathbf{\widetilde{D}^{-1/2}_{(k)}} \mathbf{\tilde{A}_{(k)}} \mathbf{\widetilde{D}^{-1/2}_{(k)}}  \mathbf{X_{t}^{(l)}} \Theta^{(l)}_{(k)} $ where: 

\begin{equation}
     \mathbf{\tilde{A}}_{(k)}=\mathbf{I}+\mathbbm{1}\left(\mathbf{\tilde{A}}^{k} \geq 1\right)-\mathbbm{1} \left(\mathbf{\tilde{A}}^{k-1} \geq 1\right)  |  \mathbf{\tilde{A}}_{(1)} = \mathbf{A} + \mathbf{I}; \mathbf{\tilde{A}}_{(0)} = \mathbf{I}
\label{ms}
\end{equation}
In summary, where the ST-GCN model \cite{S.Yan2018} adopted by \cite{S.Gadgil2020} factorises spatio-temporal convolutions into spatial-only and temporal-only blocks, useful for modelling information flow along space and time (but limited in their capacity to model space-time interactions), our method adds in parallel a space-time convolution module connecting all nodes in a short temporal window. Importantly,  the TCN block adopted here from MS-G3D \cite{Z.Liu2020} also improves capacity for more long-range temporal modelling by aggregating multi-scale temporal convolutions, over several temporal scales. 

\begin{figure}[h]
     \centering
     \includegraphics[scale=0.19]{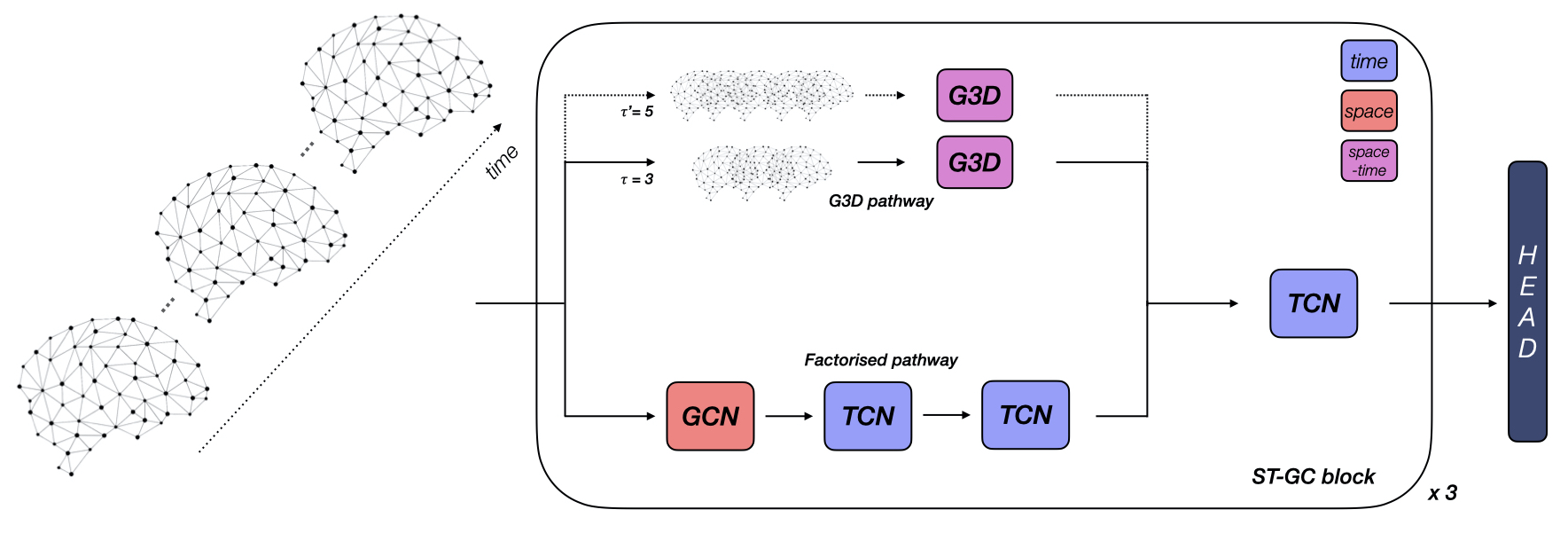}
     \caption{Architecture of our Brain-MS-G3D. A sequence of T frames is used as input of the model. Each G3D pathway processes T sub time-windows of lenght $\tau \in \left \{ 3,5 \right \}$, while the factorised pathway process the entire sequence T with spatial-only (GCN) and temporal-only (TCN) blocks. A final TCN block aggregates all pathways.}
     \label{model}
\end{figure}

\paragraph{\textbf{Architecture}}
Network implementation (illustrated in figure \ref{model}) follows the design principles of MS-G3D \cite{Z.Liu2020} but adapts it to the case of FC. At each training iteration, a time window of T consecutive frames is sampled at random from the entire sequence and is passed as input to the model. In the present architecture (figure \ref{model}), three ST-GC blocks were stacked with 96, 192, 384 outputs channels, respectively. In each of these blocks, two \emph{G3D pathways} convolutions (Eq (\ref{gctn_conv})) were implemented; these were passed spatio-temporal graphs formed from time-windows of consecutive frames \textit{at two temporal lengths} ($\tau \in \left \{3,5 \right \}$). In parallel, the factorised pathway applies spatial-only GCN (Eq (\ref{gcn_conv})), followed by temporal-only convolutions TCN, on the entire time-window sequence. Then, a multi-scale temporal module aggregates the two pathways, increasing long-range temporal modelling by connecting regional sub-windows. Finally, predictions are made either as classification or regression, by simply replacing a sigmoid activation layer (for classification) with a regression head for the fluid intelligence task. We refer to this architecture as \emph{Brain-MS-G3D}.

\paragraph{\textbf{Experimental Setup}}

We first trained models, the ST-GCN \cite{S.Gadgil2020} and our Brain-MS-G3D, on the 22 ROIs group-average cortical parcellation used by \cite{S.Gadgil2020}. Then both models were also compared on the subject-specific node time series derived from dual-regressed group-ICA at different dimensionalities (15 to 200 nodes).
Following the approach used in \cite{S.Gadgil2020},  we benchmark the proposed MS-G3D \cite{Z.Liu2020} model against the ST-GCN \cite{S.Yan2018} used in \cite{S.Gadgil2020}, by training models using 5-fold cross-validation. In both cases models were trained on different length time-windows T, from the entire 1200 time-points sequence. 
At each training iteration, a temporal window was randomly extracted from the entire sequence to be processed by the model. At inference time, predictions from $V=64$ voters were averaged. In all cases, an Adam optimiser was used with binary cross-entropy loss (for classification) and mean squared error (MSE) for regression; learning rate was set to $1e^{-3}$ by default and with a weight decay of $1e^{-3}$. All experiments were run on a single NVIDIA TITAN RTX 24GB GPU. The batch size was maximised to fit into memory depending on the node resolution: for instance, 32 elements per batch for 200 nodes but 512 for 25 nodes.

\section*{Results}

\begin{figure}[h]
     \centering
     \includegraphics[scale=0.28]{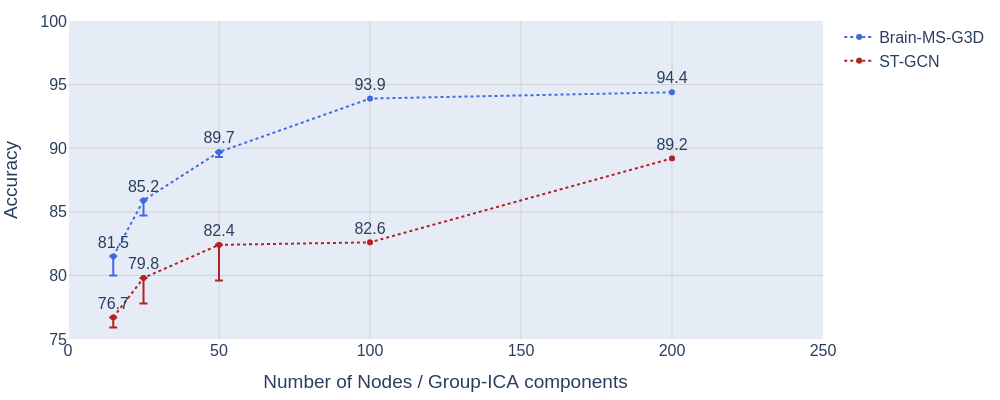}
     \caption{Comparison of sex classification results between ST-GCN and our Brain-MS-G3D on ICA subject-specific data at different resolution (15,25,50,100,200).
     }
     \label{sex_results}
\end{figure}

Results obtained are summarised in figure \ref{sex_results} and Table \ref{results}. In Table \ref{results}, we compared our method to the ST-GCN model \cite{S.Gadgil2020}, on the 22 ROIs and the ICA subject-specific node timeseries at different dimensionalities for sex classification and fluid intelligence prediction, and to results reported in a recent method which models FC dynamics using spatio-temporal attention on graphs \cite{B.Kim2021}.
In figure \ref{sex_results}, performance on sex classification is evaluated across different ICA (and group parcellation) dimensionalities, on models trained and optimised for a maximum number of 2000 iterations. Results show that our Brain-MS-G3D model, inspired by \cite{Z.Liu2020}, outperforms the ST-GCN model \cite{S.Yan2018}, used in \cite{S.Gadgil2020}, at every dimensionality, with an average increase in performance of $+6.8\%$, and a maximum of $+11.1\%$ at dimensionality 100. In Table \ref{results}, we report results for optimised trainings with more iterations, using the best time-window lenght obtained from figure \ref{sex_results}. Our Brain-MS-G3D notably reached a 94.4\% sex classification accuracy at dimensionality 200. We also note in Table \ref{results} an increase in classification for both models when moving from the group-average parcellation (22 ROIs) to a similar ICA dimensionality (25 nodes). For sex classification, we experimented the impact of changing the length of time-windows $T$ for three node dimensionalities (15,25,50): with $ T \in \left \{ 50,75,100 \right \}$, representing respectively 4.2\%, 6.3\% and 8.3\% of the entire time sequence. Range of performance across time windows is reported through the vertical bars in figure \ref{sex_results}, where plotted values correspond to the highest accuracy for $T \in \left \{ 50,75,100 \right \}$. In the case of Brain-MS-G3D, all experiments suggest an increase in performance with larger time-windows (best results with $T=100$), while the gap reduces when spatial resolution increases. 
By comparison, the ST-GCN model demonstrates a larger variance in performance, and the 100-time window network always under-performing the others. 
As widely reported in the literature \cite{R.Kong2019,R.Kong2021,U.Pervaiz2021}, fluid intelligence was a much more difficult task to predict. As detailed in \cite{U.Pervaiz2020}, performances for fluid intelligence performance are reported as correlation values in Table \ref{results}. Our results quantitatively outperform established methodology such as MAGE and HMM on fluid intelligence prediction (see \cite{U.Pervaiz2021,D.Vidaurre2018}).

\begin{table}[h]
\centering %
\setlength{\tabcolsep}{10pt}
\begin{tabular}{lccc}
        \toprule
        \textbf{Methods} & Data & \shortstack{Sex \\ (\% acc)} & \shortstack{Fluid Intelligence \\ (corr) }\\
        \midrule
        ST-GCN \cite{S.Gadgil2020}  & 22 Group ROIs & 81.8 [83.7]  & 0.144 \\
        STAGIN-SERO \cite{B.Kim2021}  & 400 ROIs & [88.20] & N/A \\
        STAGIN-GARO \cite{B.Kim2021}  & 400 ROIs & [87.01] & N/A \\
        \midrule
        Brain-MS-G3D  & 22 Group ROIs & \textbf{84.7} & 0.269 \\
        Brain-MS-G3D & ICA-15 & 81.5 & 0.286 \\
        Brain-MS-G3D & ICA-25 & 86.1 & 0.313 \\
        ST-GCN \cite{S.Gadgil2020} & ICA-25 & 82.1 & N/A \\
        Brain-MS-G3D  & ICA-50 & 90.9 & \textbf{0.325} \\
        Brain-MS-G3D & ICA-100 & 93.9 & 0.317 \\
        Brain-MS-G3D  &  ICA-200 & \textbf{94.4} & 0.324 \\
        \bottomrule
        \\
\end{tabular}
\caption{Results of sex classification and fluid intelligence prediction on the HCP dataset. 
Results presented were optimised for a maximum of 10k iterations. In square brackets, result reported in original publications.}
\label{results}
\end{table} 

\section*{Discussion}
\begin{figure}[h]
     \centering
     \begin{subfigure}[b]{0.3\textwidth}
         \centering
         \includegraphics[width=\textwidth]{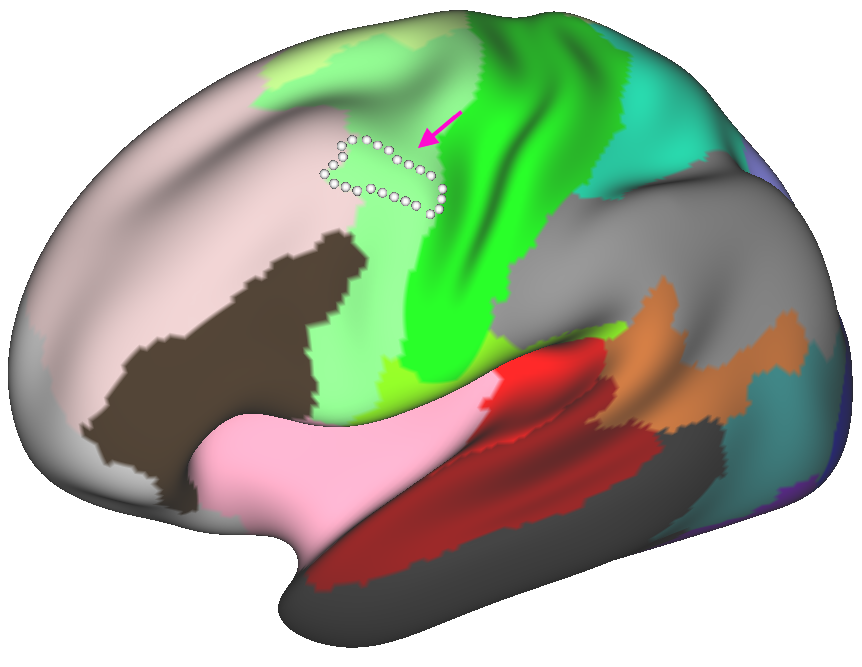}
         \caption{22 ROIs group-average parcellation derived from \cite{M.Glasser2016}}
         \label{22rois}
     \end{subfigure}
     \hfill
     \begin{subfigure}[b]{0.3\textwidth}
         \centering
         \includegraphics[width=\textwidth]{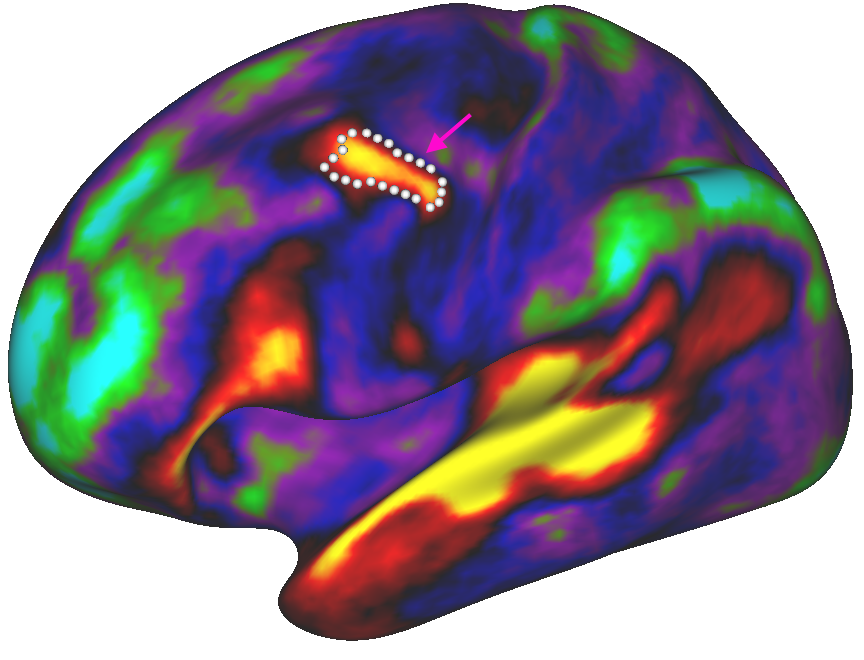}
         \caption{Group-average activation map}
         \label{group-ica}
     \end{subfigure}
     \hfill
     \begin{subfigure}[b]{0.3\textwidth}
         \centering
         \includegraphics[width=\textwidth]{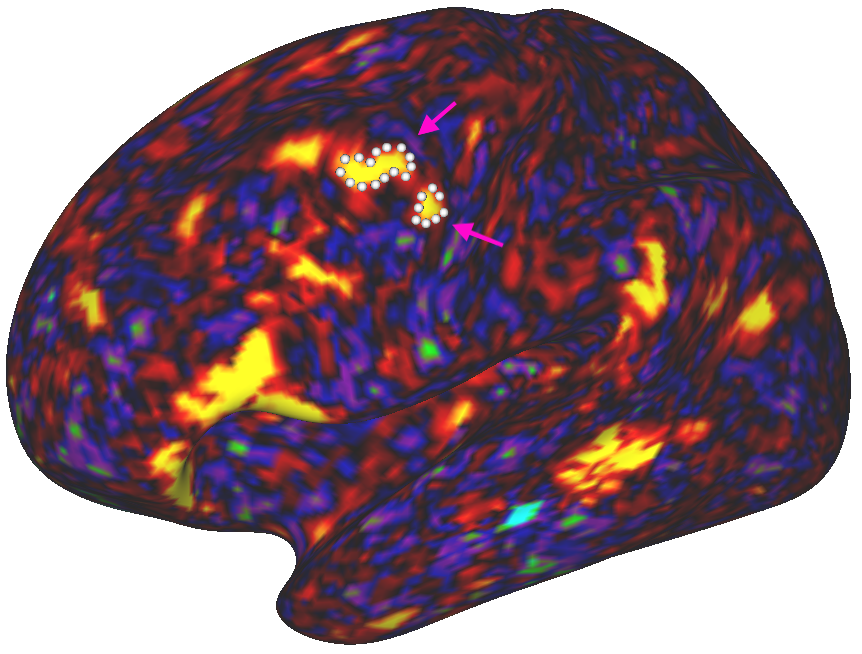}
         \caption{Subject-specific activation map}
         \label{atypical}
     \end{subfigure}
        \caption{Pink arrows highlight Area 55b, which shows topographic variability between subject. In (a), the functional connectivity of this region is averaged across the larger ROI (light green). (b) group average functional connectivity of the HCP language task contrast "Story vs. Baseline", and (c) shows atypical functional connectivity of the same language task contrast in a single subject. }
        \label{fig:parcellation}
\end{figure}

While our approach introduced a spatio-temporal modelling of functional connectivity that has proved to increase phenotype predictions, additional experiments revealed that the multi-scale operator (Eq (5)), used in \cite{Z.Liu2020}, has only a small impact on Brain-MS-G3D performances. In the case of brain connectomes, the spatial graph is by construction a weighted-undirected graph that connects all nodes together, while in skeleton analysis the spatial graph is much sparser. Thresholding correlation values in the adjacency matrix could increase the impact of spatial scaling  (Eq (5)). Additionally, in figure \ref{sex_results}, we illustrated that increasing the spatial resolution of the ICA tends to improve performances. However, the length of time sequence has also been pointed as crucial to understand the temporal hierarchy of FC \cite{D.Vidaurre2017}. We had to limit the study to a relatively small number of time frames per sequence (maximum of 128), mostly due to hardware limitations, but experiments on larger time windows with smaller ICA resolution showed dramatic increases in prediction accuracy, which should be further investigated. Finally, figure \ref{fig:parcellation} illustrates the impact of parcellation for modelling subject-variability in functional connectivity: atypical subject-specific activation might be averaged out by parcellation.  

In conclusion, this study improved behavioural and demographic predictions on the HCP dataset by introducing a state-of-the-art method for skeleton-based action recognition to explicitly modelling spatio-temporal dynamics of FC, and accounting for inter-subject variability in functional organisation through the use of dual-regressed ICA maps. Future work will explore the potential for improving performance further by increasing the spatial resolution of the model.

\section*{Acknowledgements}
Data were provided by the Human Connectome Project, WU-Minn Consortium (Principal Investigators: David Van Essen and Kamil Ugurbil; 1U54MH091657) funded by the 16 NIH Institutes and Centers that support the NIH Blueprint for Neuroscience Research; and by the McDonnell Center for Systems Neuroscience at Washington University \cite{D.VanEssen2013}. 
 
%
%
\bibliographystyle{splncs04}
\bibliography{paper35}
\end{document}